\documentclass[12pt]{article}
\usepackage{amssymb}

\usepackage{amsfonts}
\usepackage [fleqn]{amsmath}   \mathindent = 0pt
\usepackage[nohead]{geometry}
\usepackage[singlespacing]{setspace}
\usepackage{hyperref}
\usepackage{indentfirst}
\usepackage{endnotes}
\usepackage{graphicx}
\usepackage{multicol}
\usepackage{multirow}
\usepackage{appendix}
\usepackage{lscape}
\usepackage[referable]{threeparttablex}
\usepackage{tabularx}
\newcolumntype{Y}{>{\centering\arraybackslash}X}
\usepackage{booktabs}
\usepackage[capposition=top]{floatrow}
\usepackage{datetime}
\usepackage{placeins}
\newdate{date}{04}{02}{2017}
\graphicspath{ {./images/} }
\usepackage{subcaption}
\usepackage[labelsep=quad,indention=10pt]{caption}
\usepackage[labelfont=bf,list=true]{subcaption}
\DeclareMathOperator*{\argmin}{argmin}

\usepackage{amsthm}
\usepackage{color}

\usepackage[english]{babel}
\usepackage[utf8]{inputenc}

\usepackage{natbib}
\usepackage{caption}
\newtheoremstyle{query}%
{}{}
{\color{red}}
{}
{\sffamily\bfseries}{:}{12pt}
{}
\theoremstyle{query}

\usepackage{url}

\date{\vspace{-5ex}}

\begin{document}

\title{\LARGE  Big Data Information and Nowcasting: Consumption and Investment from Bank Transactions in Turkey}

\author{}

\maketitle

Ali B. Barlas (BBVA Research),
Seda Guler Mert (BBVA Research),
Berk Orkun Isa (BBVA Research)
Alvaro Ortiz (BBVA Research),
Tomasa Rodrigo (BBVA Research),
Baris Soybilgen (Bilgi University) and
Ege Yazgan (Bilgi University)

\vspace{-1cm}

\begin{center}

\vspace{1cm}

\begin{abstract}
	We use the aggregate information from individual-to-firm and firm-to-firm in Garanti BBVA Bank transactions to mimic domestic private demand. Particularly, we replicate the quarterly national accounts aggregate consumption and investment (gross fixed capital formation) and its bigger components (Machinery and Equipment and Construction) in real time for the case of Turkey. In order to validate the usefulness of the information derived from these indicators we test the nowcasting ability of both indicators to nowcast the Turkish GDP using different nowcasting models. The results are successful and confirm the usefulness of Consumption and Investment Banking transactions for nowcasting purposes. The value of the Big data information is more relevant at the beginning of the nowcasting process, when the traditional hard data information is scarce. This makes this information specially relevant for  those countries where statistical release lags are longer like the Emerging Markets.
\end{abstract}

\end{center}

\vspace{-.5cm}
\small

\linespread{1.5}

\section{Introduction}
The economists normally use the information produced by National Statistical Agencies or Central Banks (GDP, industrial production, unemployment, etc.) to assess the state of the business cycle. While this information is consistently designed to track the business cycle, it also has some shortcomings. One of the important problems is that most of the key indicators are low frequency and released with some time lag. In the case of some countries, the lag in statistical releases can be considerable.

While there is some economic information available at high frequency (i.e  stock market prices, interest rates...) it is normally related to financial conditions and expectations, which do not necessarily match the real conditions of the economy. The need to react rapidly to the changing economic conditions after the COVID-19 crisis has enhanced efforts to follow the economy in "real time" in several lines of analysis:

\begin{itemize}
	\item Focusing on alternative high frequency indicators: Some analysts have turned their attention to the more advanced released indicators such as the soft data surveys (i.e. Purchasing Manager Indexes or PMIs, Consumer confidence, etc.) and other high frequency indicators like electricity production, chain store sales released on a daily or weekly basis, respectively. 
	\item Developing higher frequency models: Central Banks have relied on the use of traditional nowcasting methods but mixing quarterly or monthly variables with higher frequency indicators (i.e. weekly, daily) to better capture the real time component information. This has been the case of the Federal Reserve of New York weekly economic index \citep{Lewis2020}, the Bundesbank Weekly Activity Index \citep{Eraslan2020} and the Central Bank of Portugal daily economic index \citep{Lourenco2020} among others.
	\item Developing New Big Data Indicators: A new stream of work \citep{Barlas2020, Carvalho2020, Chetty2020} has focused on the use of transaction data, company information or Google trends \citep{Woloszko2020} to capture the economic activity in real time. In this sense, the COVID-19 pandemic of 2020 has acted as a major stimulus for movement in this direction, and in a short space of time an entire new literature has grown using these indices.
\end{itemize}

In this working paper, we make several contributions. We extend the increasing literature of electronic payments to an emerging economy as Turkey. We use the information from individual-to-firm and firm-to-firm in Garanti BBVA\footnote{Garanti BBVA is one of the top private deposits banks operating in Turkey with its majority shareholder Banco Bilbao Vizcaya Argentaria (BBVA). BBVA is a customer-centric global financial services group operating in several countries} Bank transactions to  replicate the quarterly national accounts aggregate consumption and investment (gross fixed capital formation) and its bigger components (Machinery and Equipment and Construction). As a main contribution to the existing literature, we extend the ability of banking transactions to replicate investment  by adding firm-to-firm-transactions to the traditional individual-to-firm  used for consumptions. While the literature of Bank transactions to replicate consumption has been increasing rapidly, it is difficult to find any empirical work using Bank Data transactions to estimate investment flows. To our knowledge, only  \citet{Barlas2020} has focused in this analysis.\footnote{The real time investment information has several advantages for analysts and policymakers. First, for the case of Turkey we complete the domestic private demand by adding near a third of GDP to the consumption share during the average of the last three years. Second, investment is more volatile than consumption but with a special relevance in the source of fluctuations in Developed and specially in Emerging Economies. Third, some parts of investment as residential investment can have systemic implications for the banking and financial systems, as the 2008-2009 crisis revealed.}  

We also investigate the more efficient ways to introduce the Big Data information in alternative nowcasting models in line with other recent works \citep{Babii2014}. We use Dynamic Factor Models (DFM) and BVAR with some Machine Learning models such as Linear Lasso regressions and non linear Random Forest and Gradient Boost to test the out-of-sample nowcasting accuracy and robustness of our Big Data proxies. We find that our big data proxies significantly contribute the nowcasting performance. The contribution of the big data information appears to be more relevant at the beginning of the nowcasting process, when the traditional hard data information is relatively scarce.  

The increased availability of electronic payments data has spurred the recent literature on real time economic activity, however most of the   recent empirical works have focused on Developed economies. In the US, \citet{Barnett2016} derive an indicator-optimized augmented aggregator function over monetary and credit card services using credit card transaction volumes. This new indicator, inserted in a multivariate state space model, produces more accurate nowcasting of the GDP compared to a benchmark model. \citet{Verbaan2017} analyzes whether the use of debit card payments data improves the accuracy of the nowcast and one quarter ahead forecast of Dutch private household consumption. \citet{Baker2018} and \citet{Olafsson2018}, which use data from ﬁnancial apps to track household spending and income. \citet{Galbraith2018} generate nowcast of the Canadian GDP and retail sales using electronic payments data, including both debit card transaction and cheques clearing through the banking system. In Portugal, \citet{Duarte2017} obtain nowcast and one step ahead forecasts of Portuguese private consumption by combining data from ATM and POS terminals. In Italy, \citet{Aprigliano2017} check  retail payment data to accurately forecast Italian GDP. In Spain, \citet{Bodas2019} replicate a retail sales index through (PoS) transactions data.

The COVID-19 pandemic of 2020 has acted as a major stimulus in this direction, and in a short space of time an entire new literature has grown that uses indices derived from transaction data to track the impact of virus spread and lockdown. Again, most of the papers focused in the developed Economies. \citet{Andersen2020} presented evidence of a sharp reduction of total card spending during the early phase of the crisis in Denmark. \citet{Alexander2020}, \citet{Baker2020},  \citet{Chetty2020},  and \citet{Cox2020} focused in the response of Card transactions to COVID mobility restrictions in the USA. \citet{Bounie2020} tracked detail responses of consumer transactions in France. \citet{Chronopoulos2020} and \citet{Hacioglu2020} analyze the response in UK. Only a couple of studies have extended this empirical work to the Emerging Economies such as \citet{Barlas2020} for Turkey; \citet{Carvalho2020} for Turkey, Mexico, Colombia, Peru, Argentina; Chen et al (2020) for China. In this sense, we believe that our paper makes a significant contribution to this literature by extending it to emerging markets, deriving big data proxies for both consumption and investment flows and showing their usefulness for nowcasting purposes.

The remainder of the paper is organized as follows. In the second section, we describe the Big Data methodology to compute the private domestic demand for the case of Turkey. Particularly, we describe how to mimic national account´s consumption and investment demand from the firm-to-individuals and firm-to-firm transactions included in Garanti BBVA Big Data. Once the indicators have been developed and validated we describe their performance during the recent times in Turkey. In the third and fourth sections we describe our nowcasting methodology and check how the inclusion of the Big Data information complements and adds extra information to the traditional nowcasting Dynamic Factor Models, regularized and Machine Learning algorithms. Finally, in the last section of the paper we conclude.

\section{Consumption and Investment through a Banks' Big Data: The role of individual-to-firm and  firm-to-firm transactions} 

The recent literature of using banking transactions to replicate the household consumption has focused on the analysis of  the Point of Sales (PoS) and the  credit and debit cards data. These transaction data can be obtained from the anonymized information included in a Bank's individual database \citep{Carvalho2020,Andersen2020} or from different sources compiled by some specific companies \citep{Chetty2020}. In most of the works, individual´s consumption is estimated through the individual-to-firm payments by PoS or Card transactions exchanged by the goods and services consumed by households while some authors such as \citet{Carvalho2021} have extended this transactions to those covering the consumption of goods and services normally paid through direct money transfers such as utilities, telephone and other bills.

To extend investment demand expenditure we need to include firm-to-firm transactions received by the firms that produce fixed assets as we explained in \citet{Barlas2020}. In this case, we assume that the money transfers from firms to those firms manufacturing fixed assets are in exchange of these investment goods. In the case of Dwellings investment, we need to include t the individual-to-firm transactions reflecting the purchase of a house by households.

The definition of gross fixed capital formation by the System of National Accounts (SNA) is "the resident producers’ net acquisitions (acquisitions minus disposals) of fixed assets used in production for more than one year". The concept differentiates  between: (1) dwellings; (2) other buildings and structures, including major improvements to land; (3) machinery and equipment; (4) weapons systems; (5) cultivated biological resources, e.g. trees and livestock; (6) costs of ownership transfer on non-produced assets, like land, contracts, leases and licenses; (7) R\&D; (8) mineral exploration and evaluation; (9) computer software and databases and (10) entertainment, literary or artistic originals. Finally, these categories are grouped into three big components: (1) Construction Investment; (2) Machinery \& Equipment; and (3) Other Investment including computer software, databases, research and development, etc. In this paper, we track the first two components, Construction and Machinery \& Equipment, which in Turkey, accounts for nearly 90\% of the total investment expenditures.

In this paper, we use the transactional database of Garanti BBVA, which includes all monetary transactions between Garanti BBVA clients including individuals and firms. In the case of consumption, we identify and filter out the credit card and debit card transactions of the individuals. In this work, we take into account inflows from individuals and firms to the firms producing fixed investment assets in Machinery Investment and Construction (nearly 90\% of the total investment investment in the Turkish economy) and assume that these transfers are in exchange of the demand for fixed capital formation. 

\subsection{Consumption through Individual-to-Firm transactions}

In order to estimate the Big Data Consumption we use the Garanti BBVA Database. During 2020, this database  depicts 395.7 million filtered transactions occurred between consumers and 1.67 million merchants spread around Turkey. The information is geo-localized and include at least one transaction data from every single city in Turkey. The aggregate nominal value of those 395.7 transactions is equivalent to 299.9 billion TL (6\% of GDP)\footnote{Garanti BBVA Big Data database recorded roughly 469,000 daily consumption transactions credit cards on average in 2019, whereas median number of daily transactions was 478,000 in the same year. In 2020, average daily number of transactions decrease to 373,000 with  median  of 409,000 transactions}. 

In order to compute the our Big Data Consumption Index, we include individual-to-firm transactions including both goods and services sales. The Data extraction process for these components involve a detailed processing and filtering procedure. First, we only collects credit card and debit card transactions data from certain channels on daily basis, namely (i) physical transactions occurred at points-of-sale, (ii) e-commerce and (iii) Mail/Telephone orders. Moreover, this filtering method only includes transactions inside the 81 provinces in Turkey. 

Once the information is extracted, the data is filtered to correct for outliers and noise transactional data. Then we group the  data  by their Merchant Category Codes (MCC) to classify the information into the corresponding sectors (goods or services) and sub-sectors (airlines, restaurants, tourism, etc.). We calculate year-on-year nominal growth rates of (i) total sales transactions volume of goods, and (ii) total sales transactions volume of services. Finally we compute the Real Big Data Consumption index by deflating the nominal series with the Retail Sales deflator and the Services Sub-Item of Consumer Price Index (CPI) released by Turkish National Statistical office (TURKSTAT). 

In order to limit the sample representative bias, we weight both indexes with their respective time-varying shares in TURKSTAT Total Household Consumption Data, and weighted growth rates of two indexes are combined in order to construct the aggregate Garanti BBVA Consumption Index. 

\begin{figure}[h!]
    \setlength{\belowcaptionskip}{-10pt}
	\caption{Big Data Consumption vs National Accounts}
	\subcaption{(March 2015 to March 2021,\% YoY)} \label{fig:01}
	\centering
	\includegraphics[width=1\textwidth]{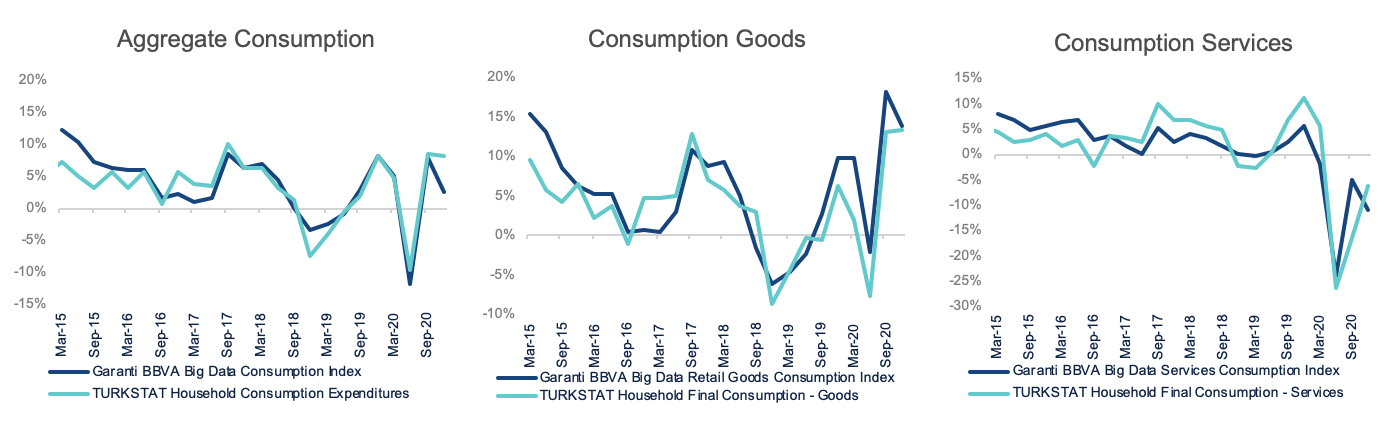}
\begin{flushleft} \footnotesize\smallskip
		\item Source: Own Elaboration and Turkstat.                             \end{flushleft}
\end{figure}

\subsection{Including Firm-to-firm Transactions to track Investment}

Our Big Data Investment Index is a synthetic indicator that aims to track investment activity based on the daily money transfers or inflows (i.e. daily purchases by means of payments) to the firms producing the fix assets. Furthermore, the key assumption is that money transfers from all the firms and individuals to those firms working in the investment-good producing sectors are in exchange of the fixed assets they produce.

The total number of firms to monitor for investment expenditures in the BBVA Big Data was nearly 367k during the year 2020. This includes 30.7k or 8\% in the Construction sector and 336.2K or 92\% in the machinery group. The aggregate number of transactions identified is 31.1 million which was worth of  1.8 trillion TL (36\% of GDP).

Table 1 shows some comparative between Garanti BBVA database and the Company Accounts Statistics 2009-2019 report published Jointly by the Central Bank of Turkey and Turkstat.\footnote{The complete survey can be accesed at http://www3.tcmb.gov.tr/sektor/2020/}. This is the more complete study published on companies in Turkey so far. Relative to this extense project the information of Garanti BBVA database would cover near 24.5\% of the sample with slightly representation of machinery manufacturing firms (25.5\%) than in Construction(19.8\%)   

\begin{table}[hbt] \label{tab:01}
	
	\centering
	\caption{Investment Firms Statistics: Garanti BBVA vs Central Bank of Turkey (CBRT) \& Turkstats Survey}
	
	\begin{tabular}{|l|c c c|c c c|}
		\hline
		& & Garanti BBVA & & & CBRT-Turkstat &  \\
		
		Variable & Tot. & Machinery & Constr. & 
		Tot. & Machinery & Constr.\\ [0.5ex] 
		\hline
		
		Transactions(000s) & 24.6 & 22.3 & 2.3 & & &  \\
		Amount(US Bn) & 308 & 280 & 28 & 440 & 257 & 183   \\ 
		Firms(000s) & 179.7 & 156.5 & 23.2 & 730.2 & 614.4 & 115.8 \\
		Firms(\% CBRT) & 24.6 & 25.5 & 19.8 &  &  &  \\
		\hline
		
	\end{tabular}
	\begin{tablenotes}[flushleft]
	    \footnotesize\smallskip
		\item Source: Garanti Bank and CBRT- Turkstat Survey.      
\end{tablenotes} 
\end{table}

In order to maintain robustness and data quality, these inflows are subject to a set of criteria . The daily money transfers data include planned installment payments, one-off payments, regular daily purchases and rest of transactions in which the counter party is not the firm that receives the payment (i.e. the transaction is not an internal transfer of funds). These inflows are filtered according to mentioned set of criteria, which consists of several rules: (i) The firm must be an active entity, that is, firms that went out of business are automatically filtered out, (ii) city of the debtor party must be identified, (iii) sector of the firm (agriculture, machinery, construction, etc.) must be identified.

The inflows data fulfilling these criteria are then pooled and sorted by their sectors according to NACE Rev. 2 classification. Investment good producing sectors are then selected based on the sectoral distribution of Gross Fixed Capital Formation in the Use Table (2012) released by TURKSTAT and the mapping between Capital Goods and NACE codes in Main Industrial Groupings (MIGs) classification.  Based on this classification, the sectoral inflows are grouped into Machinery and Construction sub-components which corresponds to around 90\% Investment expenditures in the official series.

In order to compute real investment we follow a similar process to the one explained in the case of Big Data Consumption Index. Once we estimate the nominal values for Machinery and Construction investment proxies, we deflate their annual growth rates with the Domestic Producer Price Index (D-PPI) to obtain the real growth rates. We also deflate the nominal series with the corresponding producer price deflators, but in the absence of a complete set of D-PPI individual deflators for all of the components and the fact that the results do not change significantly, we opted to use the general D-PPI for all of the components. The real year-on-year growth rates of two designated sectors, namely machinery and construction, are derived for the purpose of generating Garanti BBVA Big Data Construction Investment and Garanti BBVA Big Data Machinery Investment indexes. Last but not least, two investment indexes are once again weighted in consonance with their respective shares in TURKSTAT Gross Fixed Capital Formation data and combined in the interest of generating Garanti BBVA Big Data Investment Index.

\begin{figure}[h!]
	\setlength{\belowcaptionskip}{-10pt}
	\caption{Big Data and National Account Investment Aggregates} \subcaption{(March 2015 to March 2021, \% YoY Nominal)} 
	\label{fig:02}
	\centering
	\includegraphics[width=1\textwidth]{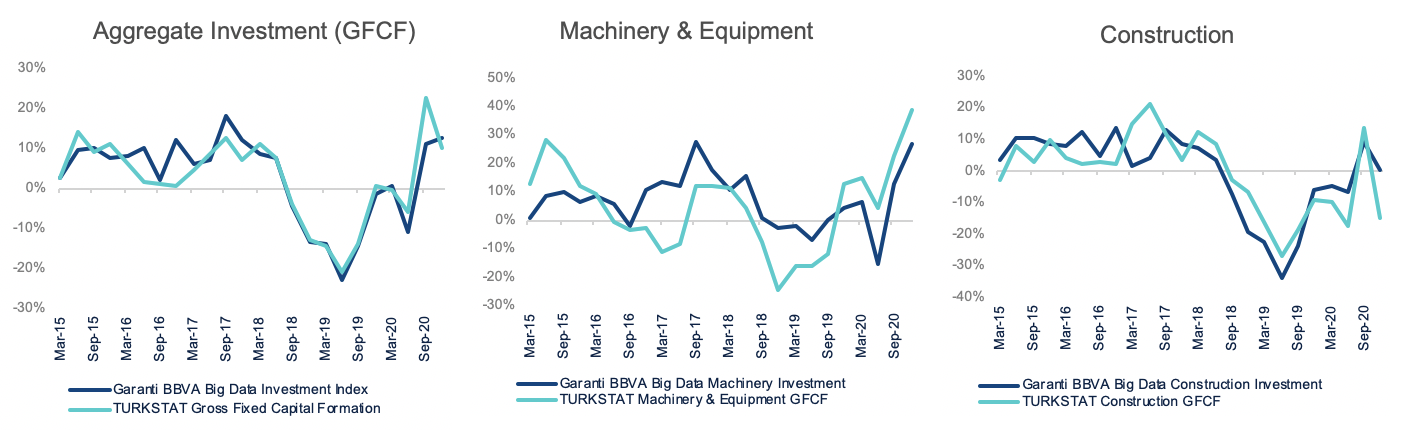}
\begin{flushleft} \footnotesize\smallskip
		\item Source: Own Elaboration and Turkstat.                             \end{flushleft}
\end{figure}

As we add the firm-to-firm money transfers to those sectors producing fixed assets and the information is geo-localized, we can estimate the evolution of fixed assets components and their geographical distribution.Thus, the Big Data Investment Indicators developed in this work present the advantage to collect information not only in real time but also in very granular way. 

Figure 3 shows the evolution of the Big Data investment assets excluding other investment. The different investment assets are grouped by two big aggregates: machinery and transport and Construction. The heat map shows the evolution of the yearly growth rate (three-month moving averages) from 2015 to 2021 (may). The darker blue colors stand for the lowest 10th percentile while the lighter blue colors represent the higher growth rates of the 90th percentile. 

\begin{figure}[h!]
	\setlength{\belowcaptionskip}{-10pt}
	\caption{Big Data Investment Sectoral HeatMap}
	\subcaption{(\% YoY Light Colours stand for positive growth rates and Dark Colours for negative rates)} \label{fig:03}
	\centering
	\includegraphics[width=1\textwidth]{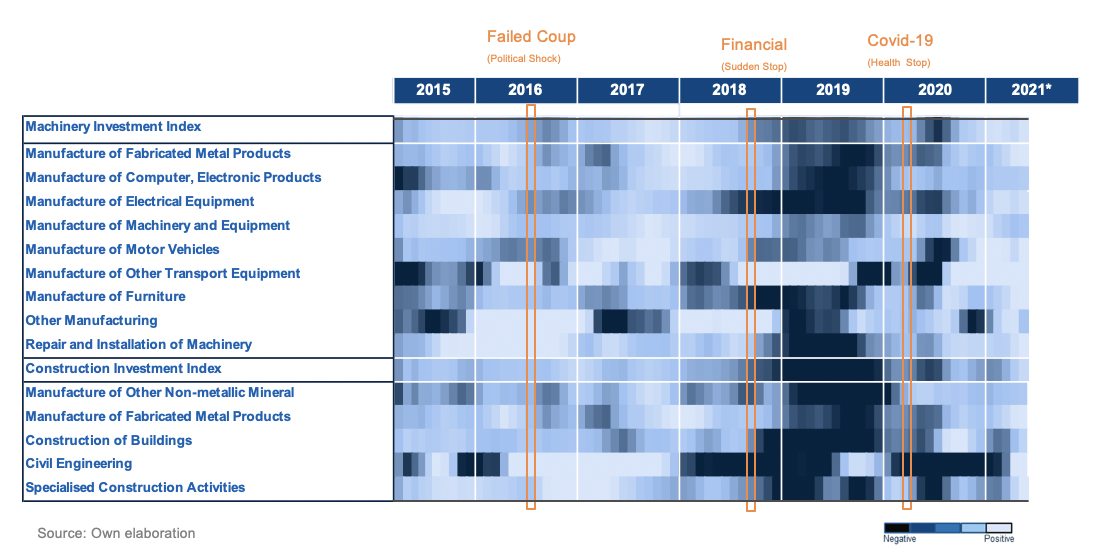}
\end{figure}

The period 2015-2021 different nature of shocks hitting the Turkish economy during this period (2015 to 2021) makes Turkey an interesting case of study to assess the impact of these investment shocks and its evolution over time on sectoral basis. In this period, three important and different shocks hit the economy (marked in orange): the failed coup of July of 2016, the financial crisis during the summer of 2018 and the recent COVID-19 shock. The response of the investment to these events has been somewhat different.

The response to the failed coup of the summer of 2016 (a political uncertainty shock) was short-lived and concentrated on some specific activities (darker colors), as shown in the non-homogeneous dark colors in the diffusion map. The negative impact was mild and mainly observed in motor vehicles and transportation fixed assets, while construction and machinery equipment was not specifically affected and recovered very rapidly once the initial political shock died out. 

The response to the summer 2018 financial crisis (a full blown currency shock) was more intense and homogeneous. After the sharp depreciation of the Turkish lira (nearly 40\%), the capital flows suddenly stopped and credits declined rapidly for most activities. The shock to investment started just after the financial crisis across the board for most of the sectors until the end of 2019 when credits started to grow after one-and-a-half years of rapid and deep de-leveraging of the sector.

The response to COVID-19 (pandemic shock) has been somewhat in the middle of the previous two shocks in terms of intensity, homogeneity and length. One relevant aspect is that the temporary shock to investment look to be related with the sudden but short sudden stop in the Global Value Changes(GVC) in which Turkey is more active as metals and Automobiles (i.e. transportation). The bigger and with longer lasting effects are concentrated in the investment of big Civil Engineering projects which could be more intensively affected by the uncertainty of the Pandemic.

Another important advantage of the Big Data financial transactions is that all the transactions are geo-localized. This is important because allows to track the investment activities not only by asset but also geographically. 

\begin{figure}[h!]
		\setlength{\belowcaptionskip}{-10pt}
	\caption{Big Data Regional Investment Maps}
	\subcaption{(\% YoY Light Colours stand for positive growth rates and Dark Colours for negative rates)} \label{fig:04}
	\centering
	\includegraphics[width=1\textwidth]{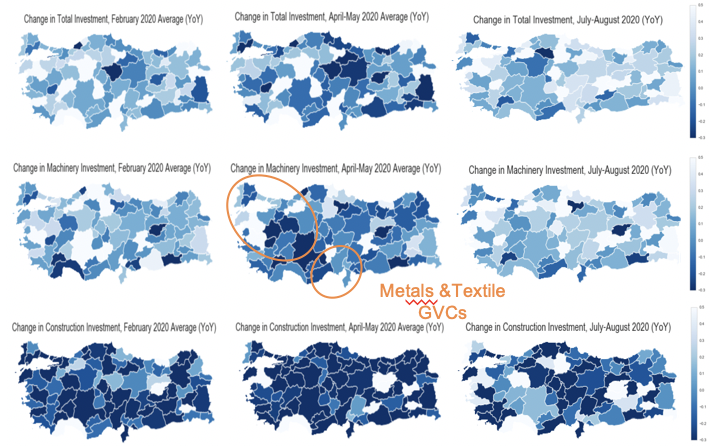}
	\begin{flushleft} \footnotesize\smallskip
		\item Source: Own Elaboration.                             \end{flushleft}
\end{figure}

The post Pandemic shock to the GVC is a good example of how this information can help us to identify different types of shocks.  Figure 4 shows how the COVID-19 shock has affected the different investment assets (Total, Machinery \& Equipment and Construction) on a geographical basis. The Heat maps show the yearly growth rates immediately before the mobility restrictions started to be imposed in Europe (February 2020), the period including most of the mobility restrictions (April-May 2020) and the ease of mobility restrictions and recovery (July-August 2020). 

The maps confirm some of our observations from the evolution of the fixed assets performance showing that the negative effects on investment by the COVID-19 shock have not been as homogeneous as the 2018 financial shock either in sectoral nor in geographical terms. The key reason for this is that the machinery investment response has been more differentiated and short lived after the COVID-19 shock. An important result is that big cities such as Istanbul or Ankara were not specially affected compared to other regions. Meanwhile, as we move from the east to the west of the country, we observe darker blues representative of contractions. This is not strange at all, as it is precisely where the manufacturing industry is mainly located \citet{Akcigit2019} and is consistent with the well-known regional east-west dualism in Turkey \citet{Gezici2017}. The maps also suggest that provinces specialized in products such as metal and electrical equipment (Central-West), mostly related to the automotive and durable consumer goods industries in the Global Value Chain (GVC), experienced the sharpest temporal declines compared to the textile industry in the GVC, which has an important presence also in the central-south of the country. The fact that the shock has been less permanent after COVID-19 is also relevant in geographical spillovers, as these GVCs are important sources of spillovers to other industries. 

The response of construction investment has been more homogeneous, but it is also recovering faster so far than the 2018 financial crisis. Facing a more negative performance than the machinery segment in the pre-COVID-19 period (as the construction sector was experiencing some de-leveraging led by the previous financial crisis), the initial response was homogeneous and amplified the already weak situation. However, the situation from June onward started to improve, at least in the big cities and the coast, but with some dark blue areas in the middle of the country. Whether this is the result of different allocation of credit or a different response of the macro-prudential policies implemented by policymakers during COVID-19 is beyond this research.

\subsection{Data included in the Nowcast Exercise}

The data set used in our analysis include several variables at different frequencies. We use quarterly chain-linked GDP series from TURKSTAT, transformed in year-on-year growth rates, as the main target variable in our nowcasting exercises. In terms of explanatory variables, we employ a total of 13 series representing broad categories of the Turkish economic activity including hard data as the labour market, manufacturing and  international trade combined with financial data, soft data represented by corporate surveys and big data activity indices including consumption and investment

We present the list of series used in Table 2 together with the vintages used in the nowcasting exercises, their frequencies and applied transformations. Since our main interest is to nowcast year-on-year GDP growth rate, for all variables, we utilized raw, non-seasonally and calendar adjusted, form of the series. All series are in real-form except for (nominal) total loan growth trend, which is deflated by the headline CPI index released by TURKSTAT. 

The data used in the nowcasting exercise is unbalanced in terms of different length of the time series and the release lag structures. This will be  addressed with different techniques depending on the models that we employ in the following sections. All variables except for the big data information are publicly available.

\begin{table}[hbt]
	
	\centering
	\caption{Detail of Variables Included in the Nowcasting Models}
	
	\begin{tabular}{ l|c c c c }
		\hline
		Variable & Type & Frequency & StartDate &Transformation \\ [0.5ex] 
		\hline
		
		GDP & Hard & Quarterly & 2003 & YoY Growth \\
		Industrial Production & Hard &  Monthly &2006 &YoY Growth   \\ 
		Auto Imports&   Hard &  Monthly & 2006 & YoY Growth \\
		Auto Sales  & Hard  & Monthly & 2003 & YoY Growth \\
		Auto Exports & Hard & Monthly & 2006 & YoY Growth \\
		Non Metalllic Minerals & Hard & Monthly &2006 &YoY Growth \\
		Electricity Production & Hard & Daily &2003 &YoY Growth \\
		Number of Employed & Hard & Monthly & 2006 & YoY Growth \\
		NUmber of Unemployed & Hard & Monthly & 2006 & YoY Growth \\
		PMI & Soft  &Monthly &2006&  Level \\
		Real Sector Confidence & Soft &Monthly &2003 &Level \\
		Loans (Credit) & Hard &Weekly& 2006 &Ann 13-week Growth \\
		Big Data Consumption  & Hard &Daily &2015 &YoY Growth  \\
		Big Data Investment & Hard & Daily & 2015 & YoY Growth \\ \hline
		
	\end{tabular}
	\begin{tablenotes}[flushleft]
	\footnotesize
		\item Source: Own Elaboration
\end{tablenotes}
\end{table}

\section{Nowcasting Methodology}

In this section we describe our nowcasting methodology. We use linear and nonlinear bridge equation models, dynamic factor models (DFM), and Bayesian vector autoregressive models (BVAR) to nowcast year over year (YoY) GDP growth rates. To build linear and nonlinear bridge equation models, we follow a similar approach to \citet{Soybilgen2021}. The biggest difference between our approach and \citet{Soybilgen2021} is that we insert variables directly into bridge equations whereas \citet{Soybilgen2021} first estimate dynamic factors and then use them in bridge equation models. DFMs are estimated following \citet{Modugno2016} and \citet{Banbura2014} and BVAR is estimated following \citet{Ankargren2019}.

The DFM used in this study can deal with the missing data at the start of the dataset, but we need to have a balanced dataset to estimate bridge equation models and BVAR. As our dataset is highly unbalanced, we follow \citet{Stekhoven2012} to fill out the missing data at the beginning of the dataset.

\subsection{Bridge Equations}

Let's define \(x_{t_m}=(x_{1,t_m},x_{2,t_m},\dots,x_{n,t_m})', t_m=1,2,\dots,T_m\) as $n$ monthly standardized explanatory variables. To construct bridge equations, we convert monthly explanatory variables to quarterly ones, \(x_{t_q}=(x_{1,t_q},x_{2,t_q},\dots,x_{n,t_q})', t_q=1,2,\dots,T_q\), by taking simple averages of $x_{t_m}$. If there is any observation missing at the end of the dataset for the reference quarter(s), we fill missing observations using an autoregressive model whose lag is chosen by the Akaike Information Criterion (AIC). Then, quarterly explanatory variables and quarterly GDP growth rates are linked as follows:
\begin{equation} \label{eq1}
y_{t_q} = g(x_{t_q}) + \varepsilon_{t_q},\\
\end{equation}    
where $g()$ defines a linear or a nonlinear functional form. In this study, we estimate equation~\ref{eq1} using ordinary least squares (OLS), random forests (RF), and gradient boosted decision trees (GBM).

RF proposed by \citet{Breiman2001} is an ensemble machine learning model based on bagging (bootstrap aggregating) of decision trees. In RF, we first obtain $B$ bootstrapped training sets from original data and then fit a decision tree to each bootstrapped training set while allowing only a random sample of variables to be considered in each variable/split point for each terminal node of a decision tree. Let $b=1, \dots ,B$ denote the number of bootstrap iterations. Then following \citet{Hastie2009}, we predict QoQ GDP growth rates using RF as: 
\begin{enumerate}
	\item Obtain the bootstrapped data from the original data; 
	\item Using the bootstrapped data obtained in the previous step, estimate a regression tree, $\hat{g}_{RF}^{(b)}$, by considering just a fraction of variables, $p$, at random from $n$ variables when determining the best variable/split point for each terminal node of a decision tree;
	\item Repeat steps 1 and 2 $B$ times;
	\item Obtain predictions of GDP growth rates as $\frac{1}{B}\sum_{b=1}^{B}\hat{g}_{RF}^{(b)}(x_{t_q+h_q})$.
\end{enumerate}

GBM is another decision tree based ensemble machine learning model. The difference between GBM and RF is that GBM turns weak learners into strong learners in a sequential way instead of separately as in RF. After an initial estimate, each tree is ﬁtted to the pseudo-residual, the gradient of the cost function, of the previous estimate, and this ﬁtted tree is then used to update the current estimate according to different learning rates for each region of a decision tree. Let us define $m=1, \dots ,M$ as the number of boosting iterations, $\lambda$ as the learning parameter, and $L()$ as the loss function. Then Following \citet{Friedman2002} and \citet{Hastie2009}, we predict QoQ GDP growth rates using GBM as:
\begin{enumerate}
	\item Initialize $g_{GBM}^{(0)}(\boldsymbol{x})=\displaystyle\argmin_\gamma\sum_{t_q=1}^{T_q}L(y_{t_q},\gamma)$;
	\item Compute the gradient of the cost function, $ r_{t_q,m} = \displaystyle\left [ \frac{\partial L\!\left(y_{t_q},g_{GBM}^{(m-1)}(x_{t_q})\right)}{\partial g_{GBM}^{(m-1)}(x_{t_q})} \right ]$;
	\item Fit a decision tree to $r_{t_q,m}$ giving terminal regions of the decision tree, $R_{j,m}, j=1,2, \dots ,J_m$;
	\item For $j=1,2, \dots ,J_m$, compute $\gamma_{j,m}=\displaystyle\argmin_\gamma\sum_{g_{t_q} \in R_{j,m}}L\!\left(y_{t_q},g_{GBM}^{(m-1)}(x_{t_q})+\gamma\right)$;
	\item Update $g_{GBM}^{(m)}(\boldsymbol{x})=g_{GBM}^{(m-1)}(\boldsymbol{x})+\displaystyle\lambda \sum_{j=1}^{J_m}\gamma_{j,m}\mathbb{I}(\boldsymbol{x}\in R_m)$;
	\item Repeat steps  2, $\dots$, 5 $M$ times;
	\item Derive the final model $g_{GBM}(\boldsymbol{x})=g_{GBM}^{(M)}(\boldsymbol{x})$;
	\item Obtain predictions of GDP growth rates as $g_{GBM}(x_{t_q+h_q})$.
\end{enumerate}

For the linear bridge equations model, we obtain predictions of GDP growth rates as $\hat{c}+\hat{\beta} x_{t_q+h_q}$ where $\hat{c}$ and $\hat{\beta}$ are estimated OLS coefficients of equation~\ref{eq1}.

\subsection{Dynamic Factor Models}

We model the DFM whose idiosyncratic components, \(\epsilon_{i,t}\), follows an AR(1) process as:
\begin{align} 
x_{t_m} &= \Lambda f_{t_m}+\epsilon_{t_m}; \label{eq2} \\ 
\epsilon_{t_m} &= \alpha \epsilon_{t_m-1} + v_{t_m}; \quad v_{t_m}\sim i.i.d.\; \mathcal{N}(0,\sigma^{2}),
\end{align}
\noindent where \(\Lambda\) is an \( n\)x\(r\) vector containing factor loadings and \(f_{t}\), which is an \(r\)x\(1\) vector of unobserved common factors, modelled as a stationary vector autoregression process as follows:
\begin{equation} 
f_{t_m} = \varphi(L) f_{t_m-1}+ \eta_{t_m}; \quad \eta_{t_m}\sim i.i.d.\; \mathcal{N}(0,R),
\end{equation}
\noindent where \(\varphi(L)\) is an \(r\)x\(r\) lag polynomial matrix and \(\eta_{t_m}\) is an \(r\)x\(1\) vector of innovations. 

To include quarterly GDP growth rates into the model, we use the approximation of \citet{Giannone2013} and impose restrictions on the factor loadings as follows:
\begin{align} 
y_{t_m}^{Q} &= \bar{\Lambda}_Q[f_{t}^{'} f_{t-1}^{'}f_{t-2}^{'}] + \bar{\epsilon}_{t_m}^{Q}; \\
\bar{\epsilon}_{t_m}^{Q} &= \alpha^{Q} \bar{\epsilon}_{t_m-1}^{Q} + \bar{v}_{t_m}^{Q}; \quad \bar{v}_{t_m}^{Q}\sim i.i.d.\; \mathcal{N}(0,\bar{\sigma}^{2}), \label{eq3}
\end{align}
where $y_{t_m}^{Q}$ denote a partially observed monthly counterpart of GDP growth rates, in which the value of the quarterly series is assigned to the third month of the respective quarter and $\bar{\Lambda}_Q=[\Lambda_Q \Lambda_Q \Lambda_Q]$ is restricted factor loadings. We estimate the DFM by following the procedure proposed by \citet{Banbura2014}, which a modified version of the expectation-maximization (EM) algorithm for maximum likelihood estimation. After casting equations \ref{eq2}-\ref{eq3} as state space form, Kalman filter and smoother allow us to extract the common factors and generate projections for all of the variables in the model.

\subsection{Bayesian Vector Autoregressive}

Let's assume that $x_{t_m}^{QM}=(x_{t_m},x_{t_m}^{Q})$ represent both observed monthly variables, $x_{t_m}$, and unobserved monthly counterpart of GDP growth rates, $x_{t_m}^{Q}$ and $X_{t_m}^{QM}=(x_{t_m},y_{t_m}^{Q})$ represent observations. Similar to the previous section, $y_{t_m}^{Q}$ denotes a partially observed monthly counterpart of GDP growth rates that can only be observed in the third month of the respective quarter and linked its unobserved monthly counterpart as follows:
\begin{equation}
y_{t_m}^{Q} = \frac{1}{3}(x_{t_m}^{Q}+x_{t_m-1}^{Q}+x_{t_m-2}^{Q}).
\end{equation}
We assume $x_{t_m}^{QM}$ follow a VAR(p) process as:
\begin{equation}
x_{t_m} = \varphi(L) x_{t_m-1}+ u_{t_m}; \quad u_{t_m}\sim i.i.d.\; \mathcal{N}(0,\Sigma),
\end{equation}

Following \citet{schorfheide2015}, \citet{sebastian2020} and \citet{Ankargren2019}, BVAR's state-space form's transition equation, which is the companion form of the VAR(p) process, and the measurement equation are shown as respectively:
\begin{align}
z_{t_m} &= \pi + \Pi z_{t_m-1}+ \zeta_{t_m}; \quad \zeta_{t_m}\sim i.i.d.\; \mathcal{N}(0,\Omega), \\
X_{t_m} &= M_t \alpha z_{t_m}
\end{align}
where $z_{t_m} = (x_{t_m}',x_{t_m-1}',x_{t_m-p+1}')'$; $\pi$, $\Pi$, and $\eta_{t_m}$ are the corresponding companion form matrices; $M_t$ is a deterministic selection matrix; $\alpha$ is an aggregation matrix.

We use the Minnesota prior for the autoregressive VAR coefficients and an inverse Wishart prior for the error-covariance. Using a Gibbs sampler, we generate draws from the posterior distributions and simulate future trajectories of $X_t$ and calculate point forecasts of all variables. Using \citet{Ankargren2019}, BVAR is estimated using Markov Chain Monte Carlo (MCMC) and Gibbs sampling.

\section{Nowcasting Performance}

\subsection{Main Results}

In this study, we estimate our models between January 2016 and December 2020 using an expanding estimation window. We assume that each prediction is computed at the end of the month and adjust announcement lag for each variable accordingly as shown in Table A1. As vintage data is not available for Turkish macro variables readily by any statistical agencies, we use a pseudo-real time dataset that ignores historical data revisions.

In each month, we produce predictions for the current quarter. We also predict the previous quarter if GDP of the previous quarter has not been announced yet. As Turkish GDP is announced with more than two months delay, we produce five predictions for each reference quarter.

We use mean absolute errors, $\textup{MAE}^{(i)}$, to evaluate the accuracy of the $i$th nowcast produced by each model between 2016Q1 and 2020Q3 as follows:
\begin{equation}\label{eq:7a}
\textup{MAE}^{(i)} = (1/n)\sum_{t_q=2016Q1}^{2020Q3} |y_{t_q}-\hat{y}_{t_q}^{(i)}|; \quad i=1,2,...,5.
\end{equation}
where $\hat{y}_{t_q}^{(i)}$ denotes the $i$th nowcast of a model and $y_{t_q}$ represents the actual GDP growth rates.

Table 3 presents MAEs for linear/nonlinear bridge equation, BVAR and DF models as well as the benchmark AR model. As can be seen from Table 3, all models outperform the benchmark AR model in all periods. In first and second nowcasts BVAR and DFM have the highest nowcasting accuracy and bridge equation models perform poorly compared to DFM and BVAR models. Note that in first and second nowcasts, we do not have any hard data available for the reference quarter, hence most of the information content in the data relies on the big data components. As a result, joint multivariate models that fill the missing part of the dataset successfully outperform bridge equation models which rely on auxiliary equations to fill out the missing data at the end of the data set. 

Starting from third predictions, we see significant improvements both in linear and nonlinear bridge equation models. For example, all bridge equation models have lower MAEs than DFM in third nowcasts. In fourth and fifth nowcasts, LM and DFM have the highest prediction accuracy, respectively. 

Overall, there is no clear winner in terms of MAEs. DFM is the best behaved model whose MAEs is steadily decreasing when informational content is improved. BVAR results are highly volatile compared to DFM and the nowcasting performances of nonlinear bridge equation models are lackluster compared to their linear counterpart.

\begin{table}[htbp] \label{tab:03}
	\centering
	\caption{MAEs of the models for successive nowcasting horizons between 2006Q1 and 2020Q3}
	\begin{tabularx}{1\textwidth}{lYYYYYY}
		\toprule
		& AR    & DFM   & BVAR  & LM    & RF    & GBM \\
		\midrule
		1st Nowcast & 3.71  & 1.92  & 1.77  & 3.46  & 2.60  & 3.13 \\
		2nd Nowcast & 3.71  & 1.85  & 2.29  & 3.07  & 2.32  & 2.55 \\
		3rd Nowcast & 3.80  & 1.72  & 1.52  & 1.70  & 1.53  & 1.71 \\
		4th Nowcast & 3.80  & 1.58  & 1.45  & 1.42  & 1.74  & 1.83 \\
		5th Nowcast & 3.80  & 1.38  & 1.64  & 1.46  & 1.65  & 1.49 \\
		\bottomrule
	\end{tabularx}%
	\begin{tablenotes}[flushleft]
	    \footnotesize\smallskip
		\item Abbreviations: AR, the benchmark autoregressive model; DFM, the dynamic factor model; BVAR, the Bayesian vector autoregressive model; LM, the linear bridge equation model; RF, the random forest based bridge equation model; GBM, the gradient tree boosted bridge equation model.
	\end{tablenotes}
\end{table}

Turkey experienced several important economic and political downturns in this period: the failed coup in 2016; the currency shock in 2018; the COVID-19 shock in 2020. Therefore, a visual inspection of the models' predictions may yield more information about the nowcasting performance of the models. Figure 5 presents nowcasts of the models for the successive nowcasting periods. In first and second nowcasts, BVAR captures the downturn in 2016Q3, the jump in 2017Q3, and the COVID shock in 2020Q2 relatively well but overestimate the recession between 2018Q4 and 2019Q3. In first and second nowcasts, DFM can not capture volatile GDP growth rates in 2016Q3 and 2017Q3. Furthermore, DFM underestimates GDP decline in 2020Q2. In other periods, DFM nowcasts GDP growth rates quite successfully. In first and second nowcasts, both linear and nonlinear bridge equation models perform quite poorly during 2017 and in 2020Q2. However, bridge equations perform much better in 2018 and 2019. Especially, RF outperform all other models in 2018 and 2019. In third, fourth, and fifth nowcasts, all models have relatively good nowcasting performance. Even though all models predict negative growth rate in 2020Q2, GBM is the best model that captures -10.3\% GDP growth rate in 2020Q2.

\begin{figure}[h!]
\setlength{\belowcaptionskip}{-10pt}
	\caption{Nowcasts Models performance and GDP}
	\subcaption{(March 2016 to September 2020,\% YoY)} \label{fig:05}
	
	\centering
	\includegraphics[width=1\textwidth]{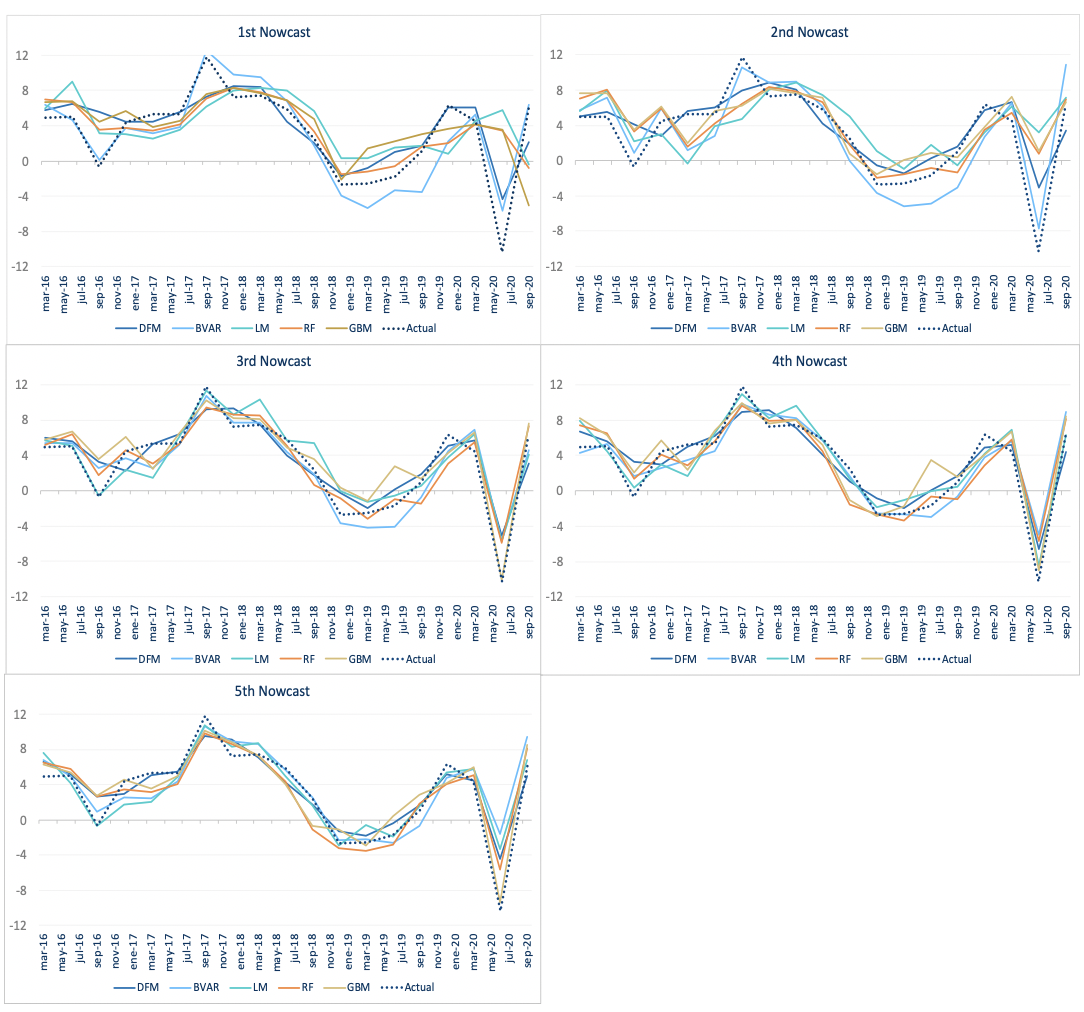}
	\floatfoot{}
    \begin{flushleft}\footnotesize\smallskip
		\item Abbreviations: DFM, the dynamic factor model; BVAR, the Bayesian vector autoregressive model; LM, the linear bridge equation model; RF, the random forest based bridge equation model; GBM, the gradient tree boosted bridge equation model.
	\end{flushleft}
\end{figure}

\subsection{Nowcast Combination}

Table 3 and Figure 5 show that there is no single best model that perform well in all periods and some models like BVAR produce very volatile nowcasts. Therefore, combining nowcasts of the models may provide better and more stable nowcasts.  We combine the prediction of each model to produce a final nowcast as follows:
\begin{equation}
\hat{Y}_{t_q}^{(i)}=\sum_{l=1}^{n}w_{t_q,l}^{(i)}\hat{y}_{t_q,l}^{(i)}, \quad l=1,2,\dots,L
\end{equation}
where \(w_{t_q,l}\) is the weight for the model \(l\) for the $i$th nowcast; $\hat{Y}_{t_q}^{(i)}$ shows the nowcast combination of the models for the $i$th nowcast; \(l=1,\dots,L\) is an index of all models. We use several types of weights to combine nowcasts in our study. These are simple weights, relative performance weights, and rank based weights.

First, we use simple averaging to calculate weights as follows: $w_{t_q,l}^{(i)}=1/L$. We also use the median forecast combination scheme. Even though the equally weighted forecast combination often outperforms sophisticated weighting techniques \citep{Clemen1989,Hendry2004,Huang2010,Stock2004}, \citet{Genre2013} and \citet{Soybilgen2018} show that advanced combination schemes may outperform equal weights in some cases.

Next, we calculate relative performance weights as:
\begin{equation}
w_{t_q,l}^{(i)}=\frac{(\text{MAE}_{t_q,l}^{(i)})^{-1}}{\sum_{l=1}^{L}(\text{MAE}_{t_q,l}^{(i)})^{-1}},
\end{equation}
where $w_{t_q,l}^{(i)}$ donates MAE of the individual model \(l\) for the $i$th nowcast calculated at time \(t_q\). We calculate MAE using the last two year nowcast performance.

We also use rank based methods to compute weights as \citet{Timmermann2006} argues that this scheme is less sensitive to outliers compared to the relative performance weight method. The rank based weights calculated as follows:
\begin{equation} \label{eq:9}
w_{t_q,l}^{(i)}=\frac{(\text{R}_{t_q,l}^{(i)})^{-1}}{\sum_{l=1}^{L}(\text{R}_{t_q,l}^{(i)})^{-1}},
\end{equation}
where $\text{R}_{t_q,l}^{(i)}$ is the rank of the model \(l\) for the $i$th nowcast calculated at time \(t_q\). Ranks are calculated according to MAEs used in the relative performance method.

Table 4 presents MAEs for various nowcast combinations and Table 5 shows MAEs for single models for the period of 2008Q2 and 2020Q3\footnote{We use previous two year performances to calculate weights.}. Simple and Median denote simple averaging and the median nowcast, respectively. RPW and Rank denote nowcast combinations calculated by relative performance weight and rank-based weights, respectively.

\begin{table}[h!] \label{tab:04}
\centering
\caption{MAEs of nowcasting combinations for successive nowcasting horizons between 2008Q2 and 2020Q3}
\begin{tabularx}{1\textwidth}{lYYYY}
\toprule
 & Simple & Median & RPW   & Rank \\
\midrule
1st Nowcast & 2.67  & 3.29  & 2.53  & 2.30 \\
2nd Nowcast & 2.03  & 2.40  & 1.95  & 1.89 \\
3rd Nowcast & 1.39  & 1.65  & 1.32  & 1.34 \\
4th Nowcast & 1.44  & 1.43  & 1.44  & 1.45 \\
5th Nowcast & 1.36  & 1.43  & 1.38  & 1.43 \\
\bottomrule
\end{tabularx}%
\begin{tablenotes}[para,flushleft]
\footnotesize
\item Abbreviations: Simple, nowcast combination using simple averaging; Median, the median nowcast; RPW, nowcast combination using the relative performance weight; Rank, nowcast combination using the rank based weight.
\end{tablenotes}
\end{table}

\begin{table}[h!] \label{tab:05}
\centering
\caption{MAEs of the individual models for successive nowcasting horizons between 2008Q2 and 2020Q3}
\begin{tabularx}{1\textwidth}{lYYYYY}
\toprule
& DFM   & BVAR  & LM    & RF    & GBM \\
\midrule
    1st Nowcast & 2.01  & 2.16  & 4.37  & 3.18  & 4.20 \\
    2nd Nowcast & 2.09  & 2.65  & 3.40  & 2.20  & 2.69 \\
    3rd Nowcast & 1.92  & 1.95  & 1.99  & 1.80  & 1.68 \\
    4th Nowcast & 1.59  & 1.57  & 1.22  & 2.06  & 1.88 \\
    5th Nowcast & 1.48  & 1.82  & 1.44  & 1.77  & 1.75 \\
\bottomrule
\end{tabularx}%
\begin{tablenotes}[para,flushleft]
\footnotesize
	\item Abbreviations: DFM, the dynamic factor model; BVAR, the Bayesian vector autoregressive model; LM, the linear bridge equation model; RF, the random forest based bridge equation model; GBM, the gradient tree boosted bridge equation model.
\end{tablenotes}
\end{table}%

Results shows that rank based weighting method has the the highest nowcast performance among all nowcast combination methods in first nowcasts, but it is still worse than DFM. In second nowcasts, all nowcast combination schemes except the median one outperform the best single model and Rank is still the best one. In third nowcasts, all nowcast combination methods including the median scheme beat the best single models. The performance difference between the most of the nowcast combination methods and the single models are substantial. In fourth and fifth nowcasts, the prediction performance of all nowcast combination schemes become very similar. 

Even though, nowcast combination schemes fail to outperform all single models in all prediction horizons, nowcast combinations outperform most of single models in many cases. Regarding the differences across weighting methods advanced nowcast combination schemes outperform simple forecast combinations in the first and second nowcasts where bridge equation models perform poorly. In third, fourth, and fifth nowcasts where all models perform quite well, both simple and advanced nowcast combination methods perform similarly.

\subsection{Variable Selection with Lasso}

If some of the variables in the dataset are not beneficial for the nowcasting performance of models, pre-selection of variables may improve results of models. In this part, we first select variables using a linear regression with L1 regularization as known as Lasso regression, and then use those selected variables in models.

In each period, we first estimate equation~\ref{eq1} with a Lasso regression and use the variables with non-zero coefficients in the main nowcasting models. Table 6 present MAEs for linear/nonlinear bridge equation, BVAR and DF models whose variables are first selected by Lasso regression. 

Comparing the results with those of Table 3, pre-selection of variables by Lasso increases nowcasting performance of the model with the exception of DFM and GBM where it leads to a decrease in nowcasting performance. As DFM is already a dimension reduction technique, an extra layer of variable selection worsens its performance. Pre-selection of variables do not seem to help the nowcasting performance of GBM either. On the contrary, selecting variables by Lasso generally improve the nowcasting performance of BVAR, LM, and RF. Overall, we see the biggest performance improvements in the linear bridge equation model. LM becomes the best model in third, fourth and fifth nowcasts with the help of Lasso regression.

\bigskip

\begin{table}[htbp] \label{tab:06}
\centering
\caption{MAEs of the models with variable pre-selection process for successive nowcasting horizons between 2006Q1 and 2020Q3}
\begin{tabularx}{1\textwidth}{lYYYYYY}
\toprule
   & AR    & DFM   & BVAR  & LM    & RF    & GBM \\
\midrule
1st Nowcast & 3.71  & 2.52  & 2.17  & 3.24  & 2.81  & 3.47 \\
2nd Nowcast & 3.71  & 2.15  & 1.45  & 2.63  & 2.07  & 2.57 \\
3rd Nowcast & 3.80  & 1.72  & 1.64  & 1.36  & 1.48  & 1.62 \\
4th Nowcast & 3.80  & 1.73  & 1.38  & 1.28  & 1.73  & 1.76 \\
5th Nowcast & 3.80  & 1.64  & 1.36  & 1.08  & 1.56  & 1.56 \\
\bottomrule
\end{tabularx}%
\begin{tablenotes}\footnotesize\smallskip
		\item Abbreviations: AR, the benchmark autoregressive model; DFM, the dynamic factor model; BVAR, the Bayesian vector autoregressive model; LM, the linear bridge equation model; RF, the random forest based bridge equation model; GBM, the gradient tree boosted bridge equation model.
\end{tablenotes}
\end{table}%

This section can be concluded that DFM appears to be the best behaved model whose MAEs is steadily decreasing when informational content is improved. It is also fairly successful across different time periods but it experiences some difficulties in capturing volatile growth periods. Nowcast combinations outperform most of single models in many cases.  Pre-selection of variables by Lasso does not improve the performance DFM. LM appears to be the best model after pre-selection.

\section{The impact of Big Data on Nowcasting}

In the previous section we have analyzed the general nowcasting performance without considering any reference to the characteristics of the variables included in the data set. In this section we will consider the role of big data variables in nowcasting performance.  

From the previous section we know that Lasso regression has helped to improve the nowcasting performance. We first look at that how many times a specific variable is selected by Lasso regression. We run Lasso regression 60 times from 2006M01 to 2012M12. Table 7 shows the percentage of periods a variable is chosen. Interestingly, variables related with car industry are never or rarely selected. These variables are usually very volatile. Industrial production series, real sector confidence index, PMI and total loans appear to be as the most frequently chosen variables. The big data variables are chosen more than 50\% of the time.

\begin{table}[htbp] \label{tab:07}
	\centering
	\caption{Percentage of periods when a variable is chosen by Lasso regression}
	\begin{tabular}{|l|c|}
		\toprule
		Name  & \multicolumn{1}{l|}{Selection Ratio} \\
		IP    & 100.0\% \\
		Car Imports & 0.0\% \\
		Ind. Production Non-Metallic Minerals & 98.3\% \\
		Car Total Sales & 1.7\% \\
		Electricity Demand & 48.3\% \\
		Number of Employed & 8.3\% \\
		Number of Unemployed & 15.0\% \\
		Car Exports & 0.0\% \\
		PMI   & 98.3\% \\
		Total Loans 13week & 83.3\% \\
		Real Sector Confidence Index & 100.0\% \\
		Big Data Consumption & 55.0\% \\
		Big Data Investment & 68.3\% \\
		\bottomrule
	\end{tabular}%
	\label{tab:addlabel}%
	\begin{tablenotes}\footnotesize\smallskip
	\centering
		\item Source: Turkstat, Markitt, OSD and Own Elaboration
\end{tablenotes}
\end{table}

\begin{figure}[h!]
	\caption{Big Data Investment and Consumption variables selection by Lasso Regression} \label{fig:06}
	\centering
	\includegraphics[width=1\textwidth]{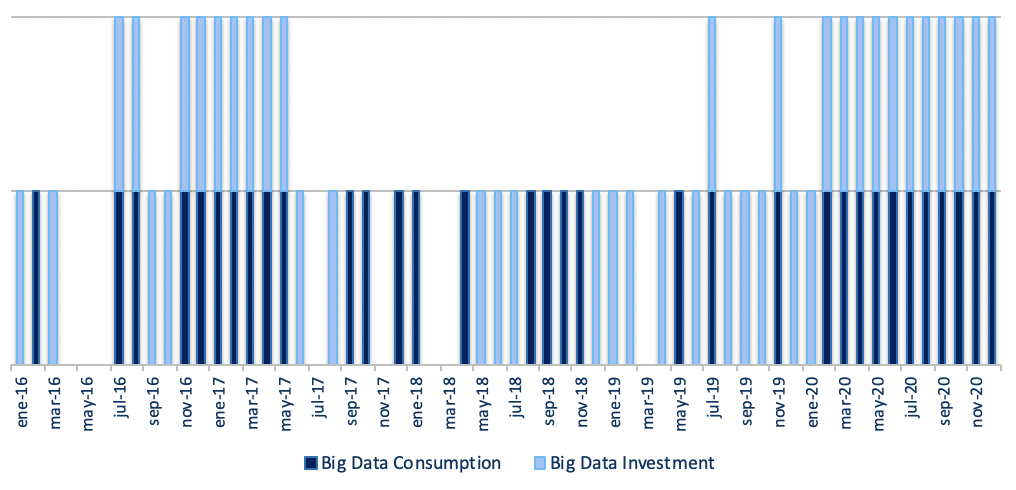}
	\floatfoot{}
	\begin{flushleft}\footnotesize\smallskip
		\item Source: Own Elaboration
\end{flushleft}
\end{figure}

In Figure~\ref{fig:06}, we illustrate the dates (between 2016M01 and 2012M12) in which the big data variables (consumption and/or investment) are chosen. Figure~\ref{fig:06} shows that most of the time Lasso regression chooses at least one big data variable. Interestingly starting from the COVID-19 shock, Lasso regression starts to pick both of the big data variables.

\subsection{The big data impact on Nowcast Accuracy}

We know analyze how our big data consumption and investment variables are beneficial for nowcasting GDP in a timely manner. To accomplish this task, first we present MAE differences (MAED) as follows:

\begin{equation}\label{eq:7b}
\textup{MAED}^{(i)} = \textup{MAE}^{(i)}-\textup{MAE}_{RD}^{(i)}; \quad i=1,2,...,5.
\end{equation}

where $\textup{MAE}_{RD}$ represents MAE of the models without big data. In other words, we estimate models with the full dataset and models with a reduced dataset that does not include big data variables. Then, we compare their nowcasting performances.

Table 8 shows MAED of the models. According to Table 8, inclusion of big data variables increase the nowcasting performance of all models in first nowcasts. The nowcast performance improvement is especially large for BVAR and bridge equation models. The big data variables provide significant improvement to LM and RF in second nowcasts, but not to DFM. Starting from third nowcasts, big data variables do not seem to exert any effects on the performance of the models in a significant way. Note that the availability of hard data concentrates usually to the third and fourth nowcast quarters. Hence in the first two nowcast quarters big data plays a big role. However when hard data become relatively abundant they become less important.

\begin{table}[htbp] \label{tab:08}
\centering
\caption{MAEDs of the models for successive nowcasting horizons between 2006Q1 and 2020Q3}
\begin{tabularx}{1\textwidth}{lYYYYY}
\toprule
& DFM   & BVAR  & LM    & RF    & GBM \\
\midrule
1st Nowcast & 0.09  & 0.57  & 0.39  & 0.51  & 0.28 \\
2nd Nowcast & 0.09  & -0.60 & 0.26  & 0.22  & 0.03 \\
3rd Nowcast & 0.07  & -0.13 & 0.01  & 0.12  & -0.04 \\
4th Nowcast & 0.06  & 0.11  & 0.00  & 0.02  & -0.29 \\
5th Nowcast & 0.05  & -0.01 & 0.06  & -0.20 & 0.01 \\
\bottomrule
\end{tabularx}%
\begin{tablenotes}[para,flushleft]
\footnotesize\smallskip
\item Abbreviations: DFM, the dynamic factor model; BVAR, the Bayesian vector autoregressive model; LM, the linear bridge equation model; RF, the random forest based bridge equation model; GBM, the gradient tree boosted bridge equation model.
\end{tablenotes}
\end{table}%

To further show the effect of big data variables on the nowcasting performance of the models, we run the models on daily basis. We assume that big data variables are released daily but other variables are announced at a specific date as shown in Table A1. For the sake of simplicity, we assume that each month consists of 30 days and calculate nowcasts for the reference quarter for 150 days until GDP is announced.

\begin{figure}[h!]
\caption{Daily MAEs of equally weighted nowcast combinations between 2006Q1 and 2020Q3} \label{fig:07}
\centering
\includegraphics[width=1\textwidth]{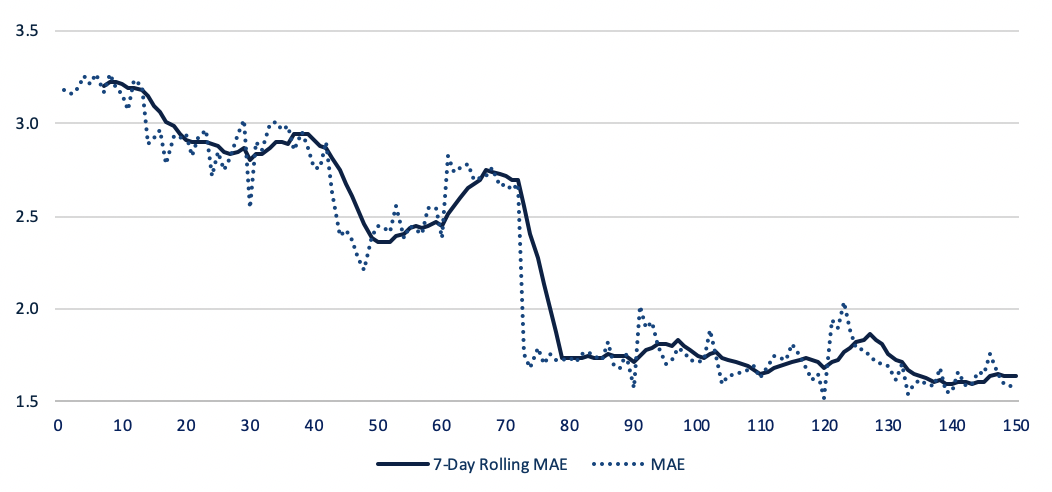}
\floatfoot{}
\begin{flushleft}\footnotesize\smallskip
		\item Source: Own Elaboration
\end{flushleft}
\end{figure}

Instead of showing each model individually, we take simple averages of all models' nowcasts. This step reduces the volatility in nowcasts of individual models and helps us to focus on the impact of variables on the nowcasting performance of the models. Figure~\ref{fig:07} shows the daily evaluation of MAEs during 150 days. As daily big data variables are highly volatile, we also show 7 day moving averages of MAEs. 

It appears that big data variables improve the nowcasting performance of the models for the first 45 days. Afterwards they become less important and do not contribute the nowcast performance in a significant way. This further strengthens our above conclusion that the big data variables are important when hard data are rarely available. We see the biggest improvement in the nowcasting performance of the models when the industrial production index' first value for the reference quarter is released in the 73th day.

\section{Conclusion}

In this paper we have developed Consumption and Investment Big Data indexes through individual-to-firm and firm-to firm transactions from a Big Data Bank Data Base (Garanti BBVA). 

The main purpose of the paper is to cross validate the information included in these real time indexes and investigate how this information can be introduced in an efficient way in the traditional nowcasting framework. For this purpose we have estimated different individual and nowcasting combinations of alternative models including Dynamic Factor Models, BVAR and Bridge equations including Linear and Non Linear Machine Learning specifications as Random Forest and Gradient Boosting. 

The first relevant result is that the Financial Transactions contribute to improve the traditional nowcasting models. Particularly, when some of the traditional economic indicators are never or rarely selected,  the Big Data information from financial transactions will be useful more than 50\% of the time.

A second important result is that Financial Transaction information is more relevant at the beginning of the nowcasting periods just when the hard data information is scarce.  Big Data information.It appears that big data variables improve the nowcasting performance of the models for the first 45 days. This conclusion strengthen our conclusion on big data variables are important when hard data are rarely available and that this is a key result for Emerging Markets where long lags in statistical releases are more important.

Besides the relevance of Big Data information we tested also the way to introduce the information. The result show that while some of the alternative produce good results in different nowcasting periods, the Dynamic Factor Model (DFM) appears to be the best alternative model as its MAEs is steadily decreasing when informational content is improved and looks  fairly successful across different time periods. Furthermore, nowcast combination schemes outperform most of single models in many cases.

In sum, the paper show the relevance of the real time financial transaction to enhance the traditional nowcasting models. Its relevance is remarked when the amount of hard data information is scarce and can help to capture turning points.  These are important results as the developing of these information is continuously increasing.

\newpage

\renewcommand{\thetable}{A.1}

\section*{Appendix}

\begin{table}[htbp] \label{tab:A1}
	\centering
	\caption{Announcement days and delays of the monthly variables}
	\begin{tabular}{|l|c|c|}
		\toprule
		Name  & \multicolumn{1}{p{10.4em}|}{Announcement Lag in Months} & \multicolumn{1}{p{6.15em}|}{Announcement Day} \\
		\midrule
		Industrial Production (IP)    & 2     & 13 \\
		Car Imports & 2     & 15 \\
		IP Non Metallic Minerals & 2     & 13 \\
		Car Sales & 2     & 15 \\
		Electricity Demand & 0     & 30 \\
		Number of Employed & 3     & 12 \\
		Number of Unemployed & 3     & 12 \\
		Car Exports & 2     & 15 \\
		Manufacturing PMI   & 1     & 1 \\
		Total Loans 13week & 1     & 10 \\
		Real Sector Confidence Index & 0     & 26 \\
		Big Data Consumption & 0     & Daily \\
		Big Data Investment & 0     & Daily \\
		\bottomrule
	\end{tabular}%
	\label{tab:addlabel}%
	\begin{flushleft}
\footnotesize\smallskip
\item Source: Own Elaboration through Turkstat, OSD, Markitt, CBRT and own Big Data
\end{flushleft}
\end{table}%

\FloatBarrier

\end{document}